\documentclass[12pt]{article}

\usepackage[utf8]{inputenc}
\usepackage[T1]{fontenc}
\usepackage{url,hyperref,lineno,microtype,subcaption}
\usepackage[onehalfspacing]{setspace}
\usepackage{lipsum}
\usepackage{algorithmicx}
\usepackage{algorithm}
\usepackage{algpseudocode}
\usepackage{dblfloatfix}
\usepackage{graphicx}
\usepackage{amsmath}
\usepackage{float}
\usepackage{cleveref}
\usepackage{newtxtext,newtxmath}

\usepackage[margin=1in]{geometry}

\usepackage{fancyhdr}
\hypersetup{
  pdftitle={GRADE: Grover-based Benchmarking Toolkit for Assessing Quantum Hardware},
  pdfauthor={Shay Manor, Millan Kumar, Priyank Behera, Azain Khalid, Oliver Zeng},
  pdfkeywords={Quantum computing, Grover's algorithm, benchmarking, reliability, quantum hardware}
}

\title{GRADE: Grover-based Benchmarking Toolkit for Assessing Quantum Hardware}

\author{
  Shay Manor$^{1}$,
  Millan Kumar$^{1}$,
  Priyank Behera$^{1}$,
  Azain Khalid$^{1}$,\\
  Oliver Zeng$^{2, *}$\\
  \small $^1$Department of Computer Science, Purdue University, West Lafayette, IN, USA \\
  \small $^2$School of Mechanical Engineering, Purdue University, West Lafayette, IN, USA \\
  \small $^*$Corresponding author:\texttt{zeng308@purdue.edu}
}

\date{}

\begin{document}
\maketitle
\begin{abstract}
Quantum computing holds the potential to provide speedups in solving complex problems that are currently difficult for classical computers. However, the realization of this potential is hindered by the issue of current hardware reliability, primarily due to noise and architectural imperfections. As quantum computing systems rapidly advance, there exists a need to create a generalizable benchmarking tool that can assess reliability across different hardware platforms. In this paper, we introduce GRADE (Grover-based Reliability Assessment for Device Evaluation), an open-source benchmarking toolkit to evaluate the reliability of quantum hardware using a generalized form of Grover's algorithm. 

GRADE operates by implementing Grover's algorithm to search through a variety of primitive collections that are customizable by the user, analyzing the probability distribution of the results to assess accuracy and stability. This approach aims to evaluate the hardware performance of Grover's algorithm, which is fundamental in unordered search problems, making it an ideal candidate for benchmarking purposes, as it is one of the few generalizable quantum algorithms that provide a direct speedup. Importantly, GRADE can adapt to a wide range of quantum computing platforms, ensuring it can be applied across different hardware architectures. Additionally, our work provides a scoring function for evaluating the hardware performance for multi-target Grover's implementation. Our work validates this approach across a multitude of simulators and hardware platforms from varying providers, demonstrating its adaptability to differing backend providers. 

\end{abstract}


\section{Introduction}
Quantum computing offers the potential for solving complex problems that are currently difficult for classical computers due to fundamental architectural differences. While classical computers process information using bits, where data is represented as either 1 or 0, quantum computers use quantum bits (qubits), which can exist in a superposition of both 1 and 0 simultaneously. This allows quantum computers to evaluate multiple solutions at once, making them faster than classical computers for specific problem sets. For example, the General Number Field Sieve (GNFS) is currently the fastest classical algorithm for factoring large integers, although sub-exponential in runtime, remains super-polynomial \cite{GNFS_1993}. In contrast, Shor's algorithm, a quantum algorithm, can factor large integers in polynomial time \cite{shor1999polynomial}.

The current phase of Noisy Intermediate-Scale Quantum (NISQ) era is characterized by quantum hardware that has only a moderate number of qubits, not yet capable of fault-tolerant quantum computation due to noise and decoherence. It is marked by the availability of quantum hardware that is capable of performing certain computations and toy examples but is also limited by its susceptibility to error. Additionally, current quantum hardware does not support error correction fully, whether it be during execution or post-measurement,  which are essential for the reliability and scalability of large-scale quantum hardware systems.

Hardware in the NISQ era is highly sensitive to noise and external disturbance, often caused by interactions between qubits and their environments, introducing errors in the computation. Noise in this context adds computational errors to the system, causing qubits to flip states or lose phase information, leading to incorrect computation results. Additionally, decoherence can cause qubits to lose their information due to extended processing. Due to the high error rate of current quantum hardware, the reliability of quantum computation is limited. Without effective error correction, the results of quantum algorithms can become unpredictable and unreliable as the circuit scales. 

\subsection{Project Scope}
As quantum computing hardware continues to advance, it has become increasingly important that quantum algorithms must be tested for performance. In this work, we propose a new benchmarking model, GRADE, which stands for Grover-based Benchmarking Toolkit for Assessing Quantum Hardware. 

Various quantum algorithms can be candidates for benchmarking, such as the Deutsch-Jozsa algorithm \cite{deutsch1992rapid}, Shor’s algorithm \cite{shor1999polynomial}, Grover’s algorithm \cite{grover1996fast}, and others. Among many quantum algorithms, Grover’s algorithm was selected as our candidate for benchmarking. Selecting specific algorithms as the benchmarking candidate, while it may seem trivial, must hold some value beyond existing as an algorithm that can be run on quantum hardware. The Deutsch-Jozsa algorithm, while popular and well known, does not fit the requirement, as there is yet to be a substantial application in the real world or integration with any well-known systems.

\subsection{Algorithm Selection}
For the scope of this project, Grover's search algorithm has been selected as the primary benchmarking algorithm. Grover’s algorithm aims to search a subset of items in an unordered database with a proven quadratic speedup \cite{nielsen2010quantum}. Additionally, Grover's algorithm has been used in real hybrid computational systems, such as supercomputing and quantum machine learning, showcasing the potential of Grover's on quantum hardware \cite{khanal2023supercomputing}. Similarly, Grover's algorithm has been used in hash functions, with the hybrid system tested on hardware successfully \cite{preston2022applying}. Furthermore, efficient algorithms have been developed using Grover's algorithm as a basis. 

Due to limitations presented by the NISQ era of quantum computing, it is crucial to benchmark algorithms that can be executed on hardware with limited qubit count while returning usable results. Grover's algorithm can be simplified for a small search, making it a potential candidate with a high success rate and ensuring that the results of the benchmark can be scaled to most systems. Subsequently, its fundamental role in search problems makes it a prime candidate for benchmarking purposes, combined with its simple implementation, making it accessible and feasible to benchmark. 

\subsection{Prior work} 
There have been many studies on benchmarking quantum computers, each using a different approach. By benchmarking quantum hardware, performance on specific problems can be evaluated for different quantum hardware. The Quantum Economic Development Consortium (QED-C) has benchmarked quantum computers using a volumetric benchmarking framework in their study \cite{lubinski2023application}. In the work presented by Lubinski \textit{et al.}, the benchmark was performed by simulating different error models (such as depolarizing errors) to mimic real quantum noise and assessing the behavior of real hardware. Rather than using a single algorithm like Grover’s search, their approach includes a benchmarking suite with algorithms from various categories, such as tutorial, subroutines, and functional. It should be noted that while their work included a comprehensive benchmark suite of multiple algorithms, including Grover's algorithm, performance was evaluated based on circuit depth rather than on target and search space. While informative about the hardware, it does not directly translate to performance on real search problems. Additionally, data included in the work only assessed Grover on simulators and not hardware. 

Similarly,  Fin{\v{z}}gar \textit{et al.} proposed QUARK, a framework for benchmarking quantum applications, with emphasis on its architecture—incorporating standardized data collection, metrics, and reproducibility—supports flexible evaluation of quantum hardware \cite{finvzgar2022quark}. While extensible across platforms like Qiskit and PennyLane, QUARK currently focuses on optimization problems and does not encompass all quantum application domains. Priority of Fin{\v{z}}gar's work has been placed on the challenges of creating representative benchmarks for quantum computing due to uncertainty over which algorithms and hardware will deliver quantum advantage.

\subsection{Novelty}
While many benchmark models have been presented previously, many studies have utilized diverse sets of algorithms with generalized metrics across multiple said algorithms. Our approach focuses on a unified benchmarking algorithm under a single algorithm's performance on multiple hardware and simulators. By using only one algorithm at a time, we provide a consistent basis for comparison across multiple different quantum hardware. Combined with a standardized scoring function that measures performance, the function quantitatively measures the performance of Grover's algorithm. The scoring function accounts for the cumulative success probability, uniformity among multiple targets, and suppression of non-target state probabilities. By standardizing the scoring criteria, the evaluation can be compared across multiple hardware providers. 

\section{Methods}
In this section, we propose a potential benchmarking model using Grover's algorithm. Our benchmark utilizes IBM Qiskit, an open-source quantum development package with the ability to transpile to multiple quantum hardware providers \cite{qiskit2024}.
\subsection{Core Concept}
The model focuses on how well a specific hardware performs with Grover's search, adjusted by two hyperparameters. As a standalone search algorithm, it is expected that the algorithm finds one or a set of primitives from a collection. In our model, we explore both single-primitive search and multi-primitive search, with adjustments made only to the oracle. The algorithm produces a score using the scoring function discussed in a later section. 
\subsection{User Input}
While the benchmarking algorithm encodes primitives as basic integer values as binary targets and search space, the user has the option to specify the size of the search space along with the number of targets to search for or specify a subset of the search space.

With search space as an optional hyperparameter without a specified input on search space, the algorithm defaults to the minimum search space requirement specified by the number of targets, calculated as $2^n$, where $n$ is the number of targets to search for, demonstrated in Alg. 1.

\begin{algorithm}[H]
\caption{GenerateSearchSpaceAndTargetsByNumTargets}
\begin{algorithmic}[1]
\Function{GenerateSearchSpaceAndTargets}{$\mathit{num\_targets}$}
    \State $n \gets \left\lceil \log_2\left(\mathit{num\_targets}\right) \right\rceil$
    \State $\mathit{search\_space\_size} \gets 2^n$
    \State $\mathit{search\_space} \gets [0, 1, \dotsc, \mathit{search\_space\_size} - 1]$
    \State $\mathit{targets} \gets \text{RandomlySelectUniqueElements}(\mathit{search\_space}, \mathit{num\_targets})$
    \ForAll{$t \in \mathit{targets}$}
        \State $\mathit{targets} \gets \mathit{targets} \cup \text{EncodeAsBinary}(t, n)$
    \EndFor
    \State \Return $\mathit{search\_space}, \mathit{targets}$
\EndFunction
\end{algorithmic}
\end{algorithm}

If the user chooses to specify a specific subset to search for without providing the search space, the model generates the minimum search space required to include the target subset, as demonstrated in Alg. 2.

\begin{algorithm}[H]
\caption{GenerateSearchSpaceForTargetList}
\begin{algorithmic}[1]
\Function{GenerateSearchSpaceForTargetList}{$\mathit{target\_list}$}
    \State $\mathit{max\_target} \gets \text{Max}(\mathit{target\_list})$
    \State $n \gets \left\lceil \log_2\left(\mathit{max\_target} + 1\right) \right\rceil$
    \State $\mathit{search\_space\_size} \gets 2^n$
    \State $\mathit{search\_space} \gets [0, 1, \dotsc, \mathit{search\_space\_size} - 1]$
    \State $\mathit{targets} \gets \mathit{target\_list}$
    \State \Return $\mathit{search\_space}, \mathit{targets}$
\EndFunction
\end{algorithmic}
\end{algorithm}

\subsection{Oracle Adjustments}
The oracle of Grover's algorithm has to be adjusted dynamically according to the target list to include all marked states. To search for multiple target states within a search space requires constructing an oracle that recognizes and inverts the phase of each target state, as demonstrated in Alg. 3.
\begin{algorithm}[H]
\caption{ConstructOracle}
\begin{algorithmic}[1]
\Function{ConstructOracle}{$\mathit{marked\_states},\ \mathit{num\_qubits}$}
    \State Initialize quantum circuit $\mathit{qc}$ with $\mathit{num\_qubits}$ qubits
    \ForAll{$\mathit{target} \in \mathit{marked\_states}$}
        \State $\mathit{zero\_indices} \gets \{i \mid \text{bit } i \text{ in } \mathit{target} \text{ is } '0'\}$
        \If{$\mathit{zero\_indices} \neq \emptyset$}
            \ForAll{$i \in \mathit{zero\_indices}$}
                \State Apply $X$ gate to qubit $i$ in $\mathit{qc}$
            \EndFor
        \EndIf
        \State Apply $H$ gate to qubit $\mathit{num\_qubits} - 1$ in $\mathit{qc}$
        \State Apply multi-controlled $X$ gate (\textit{mcx}) with control qubits $0$ to $\mathit{num\_qubits} - 2$ and target qubit $\mathit{num\_qubits} - 1$ in $\mathit{qc}$
        \State Apply $H$ gate to qubit $\mathit{num\_qubits} - 1$ in $\mathit{qc}$
        \If{$\mathit{zero\_indices} \neq \emptyset$}
            \ForAll{$i \in \mathit{zero\_indices}$}
                \State Apply $X$ gate to qubit $i$ in $\mathit{qc}$
            \EndFor
        \EndIf
    \EndFor
    \State $\mathit{oracle} \gets \mathit{qc}.\text{to\_gate}()$
    \State \Return $\mathit{oracle}$
\EndFunction
\end{algorithmic}
\end{algorithm}

The oracle marks all target states by inverting their phase relative to the rest of the superposed states, then sent to the original implementation of Grover's diffusion for state amplification. 

\subsection{Scoring}
It is critical to define a scoring function that quantitatively assesses the algorithm's performance on each hardware. Since the algorithm produces a probability distribution over all possible states, ideally, the probabilities of the target states are significantly higher than non-target states. 

\begin{algorithm}[H]
\caption{ComputeScore}
\begin{algorithmic}[1]
\Function{ComputeScore}{$P$, $T$, $\lambda$, $\mu$}
    \State $P_T \gets 0$
    \ForAll{$s \in T$}
        \State $P_T \gets P_T + P(s)$
    \EndFor
    \State $P_N \gets 1 - P_T$
    \State $\overline{P_T} \gets \dfrac{P_T}{|T|}$
    \State $\sigma_T \gets 0$
    \ForAll{$s \in T$}
        \State $\sigma_T \gets \sigma_T + \left( P(s) - \overline{P_T} \right)^2$
    \EndFor
    \State $\sigma_T \gets \sqrt{ \dfrac{\sigma_T}{|T|} }$
    \State $\text{Score} \gets P_T - \lambda \times \sigma_T - \mu \times P_N$
    \State \Return $\text{Score}$
\EndFunction
\end{algorithmic}
\end{algorithm}

\begin{figure}[H]
    \centering
    \includegraphics[width=\linewidth]{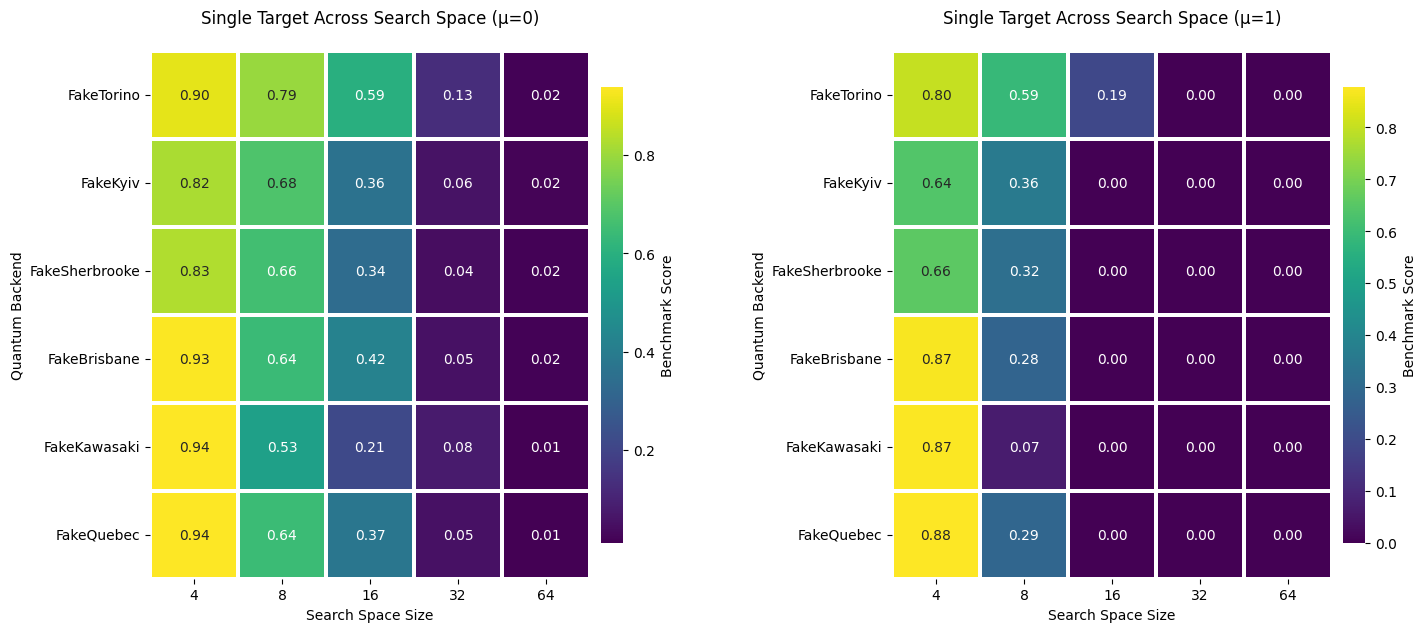}
    \caption{Heatmap comparison between different $\mu$ values, showcasing effect on GRADE scores.}
    \label{fig:1}
\end{figure}

Our scoring function, demonstrated in Alg. 4, considers the cumulative probability of the target states, the uniformity of the probabilities among the target states, and the cumulative probability of the non-target states, modeled after the scoring function for machine learning. The scoring function is defined as:
\begin{equation}
    \text{Score} = P_T - \lambda \sigma_T - \mu P_N
\end{equation}
where $P_T$ defines the cumulative probability of target states $T$, $\sigma_T$ penalizes uneven amplification among the target states, defined as:
\begin{equation}
    \sigma_T=\sqrt{\frac{1}{\vert T \vert}\sum_{s\in T}(P(s)-\overline{P_T})^2}
\end{equation}
with $\overline{P_T}$ being the mean probability of target states. $P_N$ penalizes the probability mass assigned to non-target states, defined as:
\begin{equation}
    P_N=\sum_{s\in S\char`\\T}P(s)=1-P_T
\end{equation}
$\lambda$ and $\mu$ are additional hyperparameter weighting factors that balance the importance of uniformity and non-target suppression, respectively. These penalties have been added as additional factors that can be adjusted based on benchmarking priorities, whether it is to penalize non-uniformity or penalize higher probabilities assigned to non-target states. It should be noted that when $\mu P_N$ evaluates to greater than $P_T$, the scoring function returns 0, as it implies there is a high chance of target probability being indistinguishable from the rest of the primitives in the target space, demonstrated in Fig. \ref{fig:1}. 

\section{Results}
In this section, we briefly discuss some of the results obtained from running our benchmark on a selection of simulators and quantum hardware. The current implementation of the GRADE benchmark was designed using Qiskit, combined with built-in transpilation tools for specific hardware, such as Rigetti's QuilC compiler \cite{9668289}.
\subsection{Simulators}
For simulators, we have selected fake providers from the Qiskit package provided by IBM, intended to emulate respective hardware, currently publicly accessible through IBM cloud service, with custom noise models. All simulators selected for the purpose of this work have respective hardware currently online and available through various providers at the time of this work.

\begin{figure}[H]
    \centering
    \includegraphics[width=\linewidth]{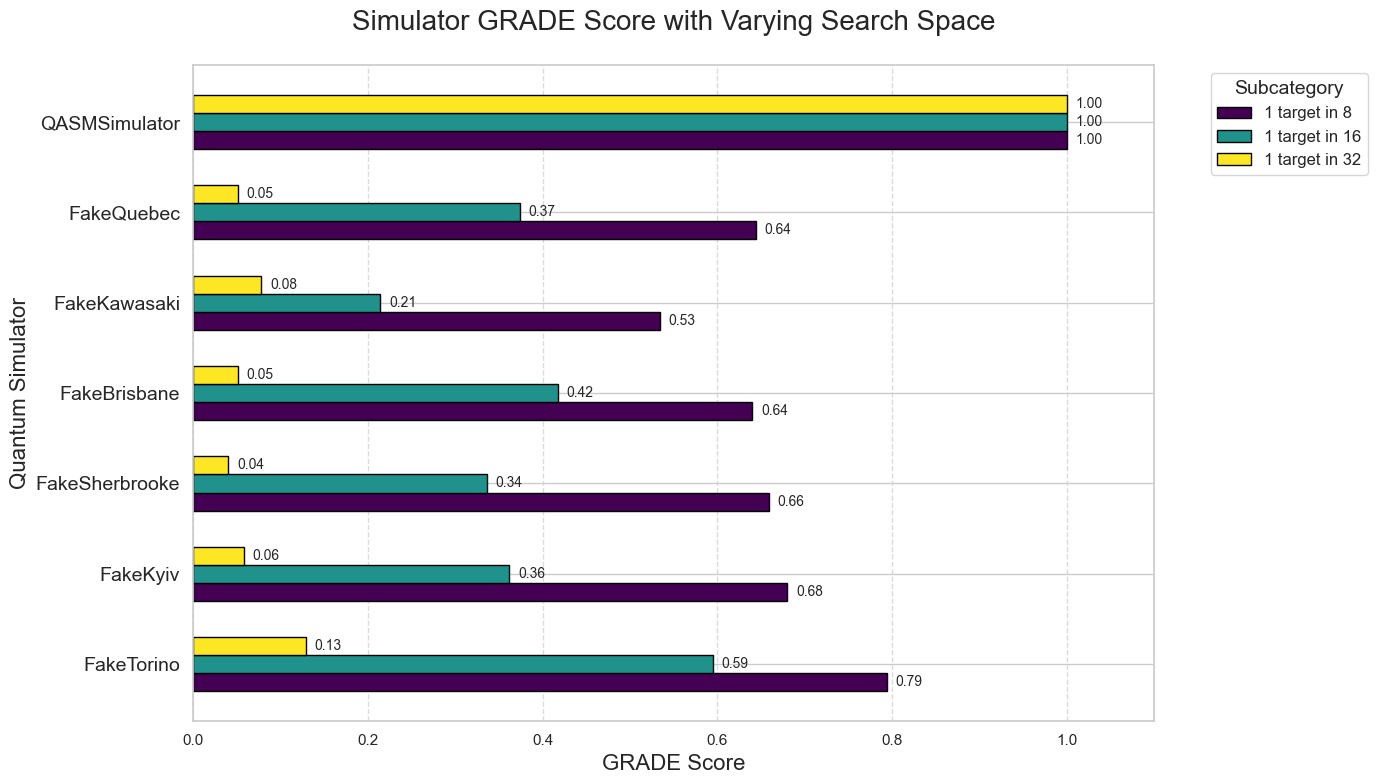}
    \caption{GRADE score of various fake backends, with $\lambda$ and $\mu$ set to 0.}
    \label{fig:2}
\end{figure}

Figure~\ref{fig:2} depicts scores collected from a selection of fake backends, searching for a single target in a varying-size search space. $\lambda$ and $\mu$ have been set to 0 for the scoring function. All benchmark scores were collected with 1000 shots, and optimization was turned off. Results indicate a decline in scoring value as the search space increases in size. It should be noted that the "QASMSimulator" is a noiseless statevector simulator and therefore will not decrease in accuracy. 

\begin{figure}[H]
    \centering
    \includegraphics[width=0.7\linewidth]{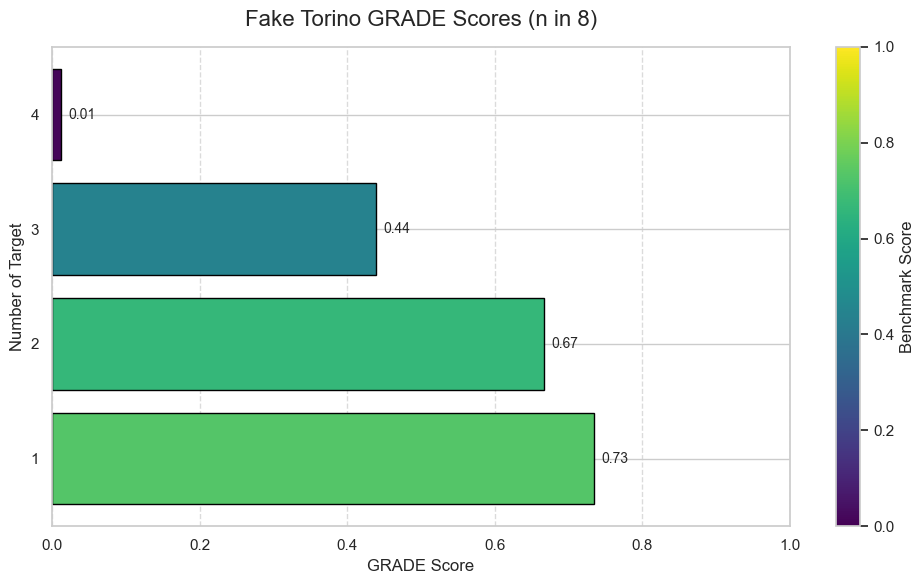}
    \caption{GRADE score report for fake Torino, benchmarked for 1 to 4 targets in search space with size 8, with $\lambda$ and $\mu$ set to 1.}
    \label{fig:3}
\end{figure}

Our benchmark is also capable of evaluating Grover's with multi-target in search space, demonstrated in Fig. \ref{fig:3}, which demonstrates 1 to 4 targets in search space with size 8, with $\lambda$ and $\mu$ set to 1.

\subsection{Hardware Testing}
As a proof of concept, we also used the GRADE benchmarking algorithm on quantum hardware providers from IonQ, Rigetti, and Quantinuum. The hardware selected for this section is currently publically accessible through Microsoft Azure. All Grover circuits run on quantum hardware were uploaded through cloud providers and returned in classical format. 

\begin{figure}[H]
    \centering
    \includegraphics[width=0.7\linewidth]{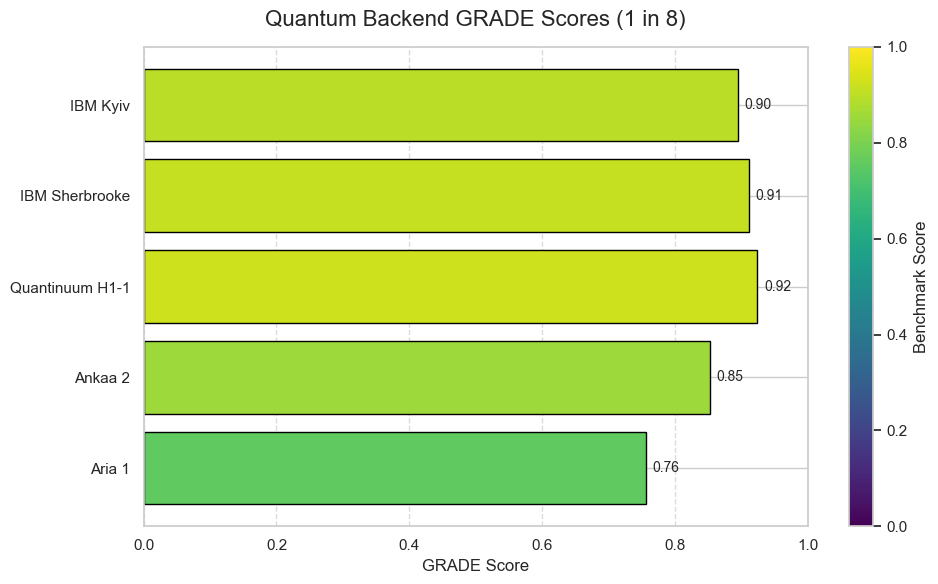}
\caption{GRADE score report for IonQ Aria 1, Rigetti Ankaa 2, Quantinuum H1-1, IBM Sherbrooke, and IBM Kyiv, benchmarked for 1 target in search space with size 8, with $\lambda$ and $\mu$ set to 1.}
    \label{fig:4}
\end{figure}

As demonstration, hardware providers were benchmarked using a search space of 8 and a target size of 1, with $\lambda$ and $\mu$ set to 1, depicted in Fig. \ref{fig:4}. As the result indicates, all hardware successfully returned a score of higher than 0.75, showcasing a usable result for searching for single primitive in a search space of size 8. While the example itself is simple, the purpose of this demonstration is not to make any evaluation or assumption about the hardware but rather to showcase the benchmark successfully being adapted to multiple hardware with comparable results. It should be noted that while the default value for $\lambda$ is 1, it does not affect the result in the cases where the search space is of size 1, as $\sigma_T$ would evaluate to 0. 

It should be noted that while we evaluated our GRADE benchmarking framework across multiple quantum hardware providers, no hardware-specific optimizations were performed. The goal of our work was to assess the general applicability and performance of Grover's algorithm as a benchmarking tool without introducing bias towards a particular hardware implementation. Consequently, the data presented are meant to reflect the algorithm's intrinsic behavior rather than the capabilities or limitations of specific hardware providers presented in this work. No claims or comparisons regarding the individual hardware platforms were made based on the results. 

\section{Conclusion and Discussion}
In this paper, we proposed a potential framework for utilizing Grover's algorithm as a benchmarking tool for evaluating quantum computer performance. By allowing the user to define specific hyperparameters that better suit to their algorithmic needs, the benchmark aims to provide flexibility in defining the search space and complexity of the problem. Our proposed scoring function accounts for the probability of the target states, the uniformity of probability distributions, and the suppression of non-target state probabilities. This add-on ensures that in scenarios involving multiple target states, the scoring function accounts for more than just the success rate. 

While the current framework provides a foundation for basic algorithmic benchmarking, it should be noted that many paths exist for future expansions. The current adaptation does not utilize any error mitigation strategies or any hardware-specific optimizations. By integrating error mitigation protocols, we can assess their efficacy and impact on the overall performance of Grover's algorithm. Additionally, extending the benchmarking framework to other quantum algorithms, such as the Quantum Approximate Optimization Algorithm (QAOA) or the Variational Quantum Eigensolver (VQE), can provide a more comprehensive toolset for quantum hardware evaluation of other common algorithms. Exploring these potential directions would provide insight into enhancing the utility and applicability of not only Grover's algorithm but also the overall algorithmic performance on various architecture platforms. With these efforts, we aim to create a baseline benchmark for identifying performance bottlenecks and optimizing quantum algorithms.

\section*{Conflict of Interest Statement}
%All financial, commercial or other relationships that might be perceived by the academic community as representing a potential conflict of interest must be disclosed. If no such relationship exists, authors will be asked to confirm the following statement: 

The authors declare that the research was conducted in the absence of any commercial or financial relationships that could be construed as a potential conflict of interest.

\section*{Author Contributions}

\textbf{Methodology and Investigation:} Shay Manor (SM), Millan Kumar (MK), Priyank Behera (PB), Azain Khalid (AK), Oliver Zeng (OZ)\\
\textbf{Data Curation:} SM, MK, PB, AK, OZ \\
\textbf{Funding Acquisition and Resources:} Donyang Li (DL) \\
\textbf{Writing – Original Draft:} SM, MK, PB, AK, OZ \\
\textbf{Writing – Review \& Editing:} SM, MK, PB, AK, OZ 

\section*{Funding}
This work was supported by the National Defense Education Program (NDEP) for Science, Technology, Engineering, and Mathematics (STEM) Education, Outreach, and Workforce Initiative Programs under Grant No. HQ0034-21-1-0014

\section*{Acknowledgments}
We would like to thank Professor Mahdi Hosseini for the support and Microsoft for providing access to quantum hardware through Azure.

\section*{Data Availability Statement}
The original contributions presented in the study are included in the article, further inquiries can be directed to the corresponding author.

\bibliographystyle{unsrt}
\bibliography{test}

\begin{thebibliography}{10}

\bibitem{GNFS_1993}
Arjen~K. Lenstra, Hendrik~W. Lenstra, Mark~S. Manasse, and John~M. Pollard.
\newblock {\em The Development of the Number Field Sieve}, volume 1554.
\newblock Springer-Verlag, Berlin, Heidelberg, 1993.

\bibitem{shor1999polynomial}
Peter~W Shor.
\newblock Polynomial-time algorithms for prime factorization and discrete logarithms on a quantum computer.
\newblock {\em SIAM review}, 41(2):303--332, 1999.

\bibitem{deutsch1992rapid}
David Deutsch and Richard Jozsa.
\newblock Rapid solution of problems by quantum computation.
\newblock {\em Proceedings of the Royal Society of London. Series A: Mathematical and Physical Sciences}, 439(1907):553--558, 1992.

\bibitem{grover1996fast}
Lov~K Grover.
\newblock A fast quantum mechanical algorithm for database search.
\newblock In {\em Proceedings of the twenty-eighth annual ACM symposium on Theory of computing}, pages 212--219, 1996.

\bibitem{nielsen2010quantum}
Michael~A Nielsen and Isaac~L Chuang.
\newblock {\em Quantum computation and quantum information}.
\newblock Cambridge university press, 2010.

\bibitem{khanal2023supercomputing}
Bikram Khanal, Javier Orduz, Pablo Rivas, and Erich Baker.
\newblock Supercomputing leverages quantum machine learning and grover’s algorithm.
\newblock {\em The Journal of Supercomputing}, 79(6):6918--6940, 2023.

\bibitem{preston2022applying}
Richard~H Preston.
\newblock Applying grover's algorithm to hash functions: a software perspective.
\newblock {\em IEEE Transactions on Quantum Engineering}, 3:1--10, 2022.

\bibitem{lubinski2023application}
Thomas Lubinski, Sonika Johri, Paul Varosy, Jeremiah Coleman, Luning Zhao, Jason Necaise, Charles~H Baldwin, Karl Mayer, and Timothy Proctor.
\newblock Application-oriented performance benchmarks for quantum computing.
\newblock {\em IEEE Transactions on Quantum Engineering}, 4:1--32, 2023.

\bibitem{finvzgar2022quark}
Jernej~Rudi Fin{\v{z}}gar, Philipp Ross, Leonhard H{\"o}lscher, Johannes Klepsch, and Andre Luckow.
\newblock Quark: A framework for quantum computing application benchmarking.
\newblock In {\em 2022 IEEE international conference on quantum computing and engineering (QCE)}, pages 226--237. IEEE, 2022.

\bibitem{qiskit2024}
Ali Javadi-Abhari, Matthew Treinish, Kevin Krsulich, Christopher~J. Wood, Jake Lishman, Julien Gacon, Simon Martiel, Paul~D. Nation, Lev~S. Bishop, Andrew~W. Cross, Blake~R. Johnson, and Jay~M. Gambetta.
\newblock Quantum computing with {Q}iskit, 2024.

\bibitem{9668289}
Gokul~Subramanian Ravi, Kaitlin~N. Smith, Pranav Gokhale, and Frederic~T. Chong.
\newblock Quantum computing in the cloud: Analyzing job and machine characteristics.
\newblock In {\em 2021 IEEE International Symposium on Workload Characterization (IISWC)}, pages 39--50, 2021.

\end{thebibliography}

\end{document}